\documentclass[10pt]{IEEEtran}
\usepackage{amsmath}
\usepackage{amsfonts}
\usepackage{amssymb}
\usepackage[ruled,vlined]{algorithm2e}
\usepackage{caption}
\usepackage{mathtools}  
\usepackage{amsthm}
\usepackage[]{algorithm2e}
\usepackage{booktabs}
\usepackage{subfigure}
\usepackage{tabulary}
\usepackage{footnote}
\usepackage[export]{adjustbox}
\usepackage{booktabs}
\usepackage[justification=centering]{caption}
\usepackage{comment}
\newtheorem{remark}{Remark}
\usepackage{cases}
\usepackage{graphicx}
\usepackage{cite}
\usepackage{multicol}
\usepackage{xcolor}
\usepackage{graphicx}
\graphicspath{{./}{figures/}}
\newcommand\numeq[1]%
  {\stackrel{\scriptscriptstyle(\mkern-1.5mu#1\mkern-1.5mu)}{=}}
\usepackage{stfloats}% <-- added
\usepackage{cuted}
\usepackage{lipsum}  

\usepackage{subfigure}

\title{Non-Primary Channel Access in IEEE 802.11 UHR: Comprehensive Analysis and Evaluation}

\author{\text{Dongyu Wei},  \text{Liu Cao}, \text{Lyutianyang Zhang}, \text{Xiangyu Gao}, \text{Hao Yin
}
\thanks{Dongyu Wei, Liu Cao, Lyutianyang Zhang, Xiangyu Gao, and Hao Yin are with the Department of Electrical \& Computer Engineering, University of Washington, Seattle, WA, USA (e-mail:\{wdy1223, liucao, lyutiz, xygao, haoyin\}@uw.edu).}}% 
\begin{document}
% Extremely high throughput (EHT) in IEEE 802.11be for future upgrade of IEEE 802.11 standard focuses on indoor and outdoor wireless local area network (WLAN) throughput maximization.  

\maketitle

\begin{abstract}
The evolution of the IEEE 802.11 standards marks a significant throughput advancement in wireless access technologies, progressively increasing bandwidth capacities from 20 MHz in the IEEE 802.11a to up to 320 MHz in the latest IEEE 802.11be (Wi-Fi 7). However, the increased bandwidth capacities may not be well exploited due to inefficient bandwidth utilization on multiple channels. This issue typically occurs when the primary channel is busy, secondary channels (also known as non-primary channels) are prevented from being utilized even if they are idle, thereby wasting the available bandwidth. This paper investigates the fundamentals of the Non-Primary Channel Access (NPCA) protocol that was defined in IEEE 802.11 Ultra-High Reliability (UHR) group to cope with the above issue. We develop a novel analytical model to assess NPCA protocol performance in terms of the average throughput and delay. Via simulation, we verify that the NPCA network outperforms the legacy network by increasing at least 50\% average throughput while reducing at least 40\% average delay. 
\end{abstract}

\section{Introduction} 
\label{introduction}
The Institute of Electrical and Electronics Engineers (IEEE) 802.11 standards represent a cornerstone in contemporary wireless access technologies, enjoying widespread application across various domains. Among these, the bandwidth capacity evolved from 20 MHz in the IEEE 802.11a to a maximum of 160 MHz in the 802.11ac/ax standards \cite{masiukiewicz2019throughput}. The latest iteration, designated as the IEEE 802.11be or Extremely High Throughput (EHT) standard, heralds a significant leap in bandwidth capacity, supporting up to 320 MHz in \cite{lopez2019ieee, hoefel2020ieee}. This enhancement introduces a paradigm shift in channel utilization strategies. Specifically, the EHT standard facilitates the allocation of secondary channels for stations (STAs) when connected to an access point (AP) supporting the full 320 MHz bandwidth. This approach mitigates the congestion of stations with smaller bandwidth capacities on the primary channel, thereby optimizing the utilization of available bandwidth across both primary and secondary channels. 

Despite these advancements, each standard version maintains a primary 20 MHz channel to preserve backward compatibility throughout the standards' evolutionary trajectory. However, mandating the primary channel in all data transmission processes may impose limitations on system efficiency because the secondary channels are prevented from being utilized when the primary channel is occupied. Despite, the innovative propositions of the EHT standard for secondary channel communications, several critical implementation aspects remain unaddressed, including the methodology for APs to access and transmit data via secondary channels when the primary channel is congested. To further increase the efficiency of the channel access methods, the IEEE 802.11 Ultra-High Reliability (UHR) group further proposes the Non-Primary Channel Access (NPCA) to take advantage of the dynamic usage of non-Primary channels \cite{IEEE80211-23/0797r0}.

%Related work

The performance determined by channel access under the IEEE 802.11 standards has been widely investigated in the existing works \cite{yin2022ieee,bianchi2000performance,1019305, 1388729,tinnirello2009refinements,ziouva2002csma,yin2020ns3,4202176,gao2023learning,cao2020performance}. Particularly, Bianchi's IEEE 802.11 DCF network model in \cite{bianchi2000performance} has been broadly used. This seminal model sets up a two-dimensional Markov chain for an assumably saturated network where the backoff action of every node is established. It is a simple yet robust model for analyzing the throughput performance of the Base Station Subsystem (BSS). Bianchi's model was further extended and refined in \cite{1019305, 1388729, tinnirello2009refinements} to improve throughput, and delay analysis of the network was also proposed in \cite{ziouva2002csma, yin2020ns3, 4202176}. In the next generation IEEE 11.be Wi-Fi standard, the performance of the model was further analyzed in \cite{Perez2019AP, zhang2023ieee, lopez2022multi}, where advancements such as bandwidth up to 320 MHz, Multi-Link Operation (MLO), and multi-band/multi-channel aggregation and operation improve the throughput and limit delay in the network. And for the upcoming IEEE 802.11bn (UHR), also known as Wi-Fi 8, anticipatory research has also been conducted towards its reliability and latency \cite{giordano2023will}.

Despite these advancements, to the best of the authors' knowledge, NPCA protocol has never been analyzed yet to help the standard designer and users realize the possible performance gain. To address this gap, we investigate the latest NPCA protocol to compare its performance gain with the legacy protocol. The main contribution of this paper can be summarized below:
\begin{itemize}
    \item We develop an analytical model for the NPCA network that is defined by IEEE 802.11 UHR group in terms of the average throughput and delay. The analytical models are validated via simulations.
    \item We compare the performance gain between NPCA and legacy protocols and demonstrate the advantages of the NPCA network over the legacy network. 
\end{itemize}

The rest of this paper is organized as follows. Section \ref{sec:sys_arc} introduces the NPCA mechanism. The proposed model in terms of the average throughput and delay in the NPCA network is presented in Section \ref{sec:sys_model}. In Section \ref{simulation}, we conduct simulations to validate our proposed model. Finally, section \ref{sec.conclusion} concludes this paper.

\section{NPCA Network Mechanism}
\label{sec:sys_arc}

% \subsection{Overview of NPCA}

\begin{figure}[tp]
    \centering
\includegraphics[width=.48\textwidth]{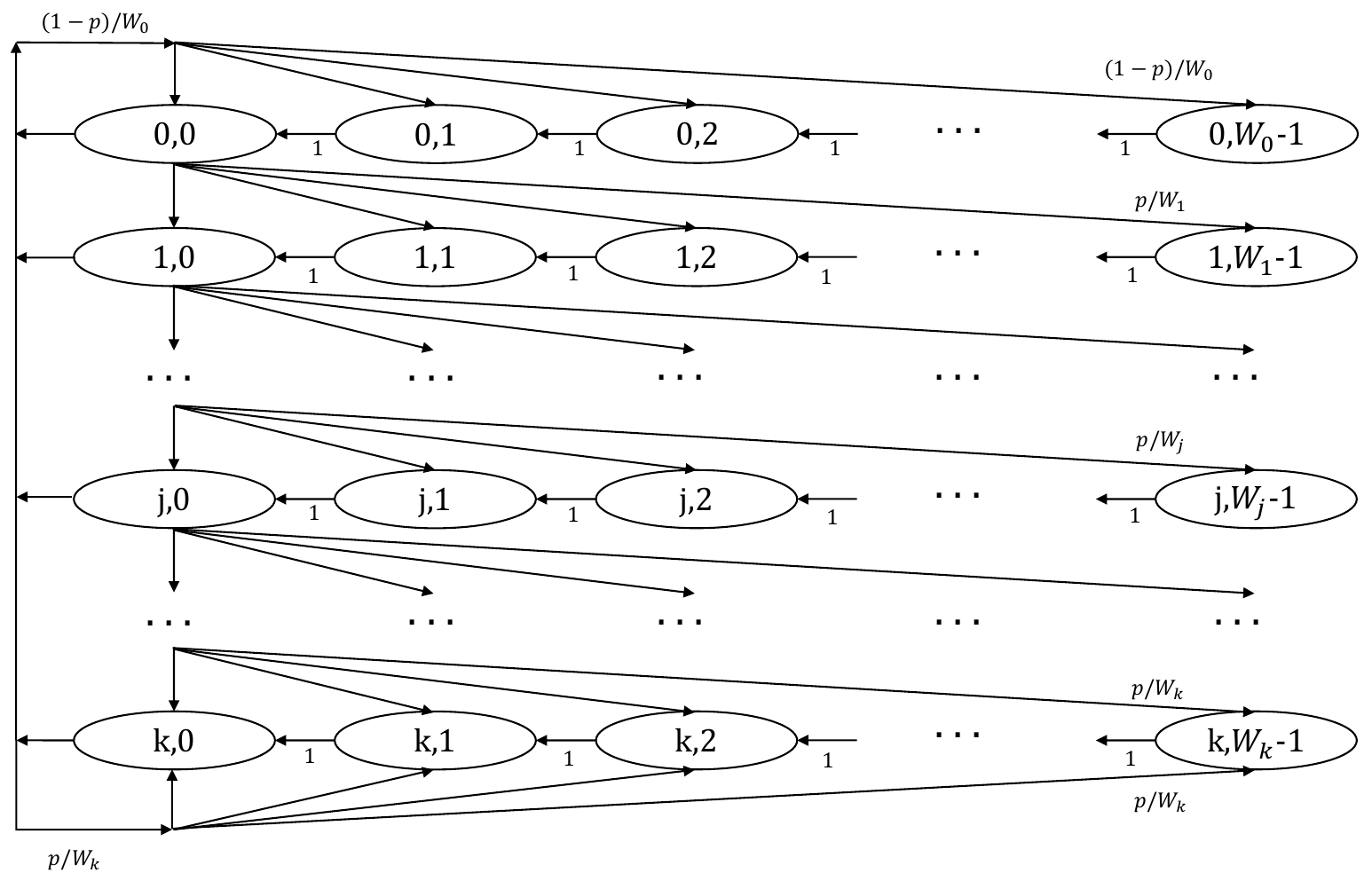}
    \caption{Markov Chain Model of Contention Window Backoff in CSMA/CA.}
    \label{fig:markov}
\end{figure}

% When a device has data to send, it first checks if the channel is free. If yes, it proceeds with a countdown based on a contention window (CW). This countdown is paused if the channel becomes busy due to other devices' transmissions.
Wireless devices that follow the IEEE 802.11 standards communicate over channels using a protocol, known as Carrier Sense Multiple Access with Collision Avoidance (CSMA/CA). 
Suppose a device wants to occupy its primary channel. It waits until the channel is sensed idle for a distributed inter-frame space (DIFS), followed by a backoff process. As Fig. \ref{fig:markov} shows, the backoff counter value is initialized by uniformly choosing an integer from the range [$0, W-1$]. Then, it is decremented by one at the end of each idle slot. Note that the backoff counter will be frozen when a packet transmission is detected on the channel and will be reactivated until the channel is sensed idle again for a DIFS period. The device occupies its primary channel and starts transmission when its backoff counter reaches zero. The contention window size $W$ is doubled after each unsuccessful transmission, up to a maximum of $k$ unsuccessful transmissions.

\begin{figure*}[t]
    \centering
\includegraphics[width=.85\textwidth]{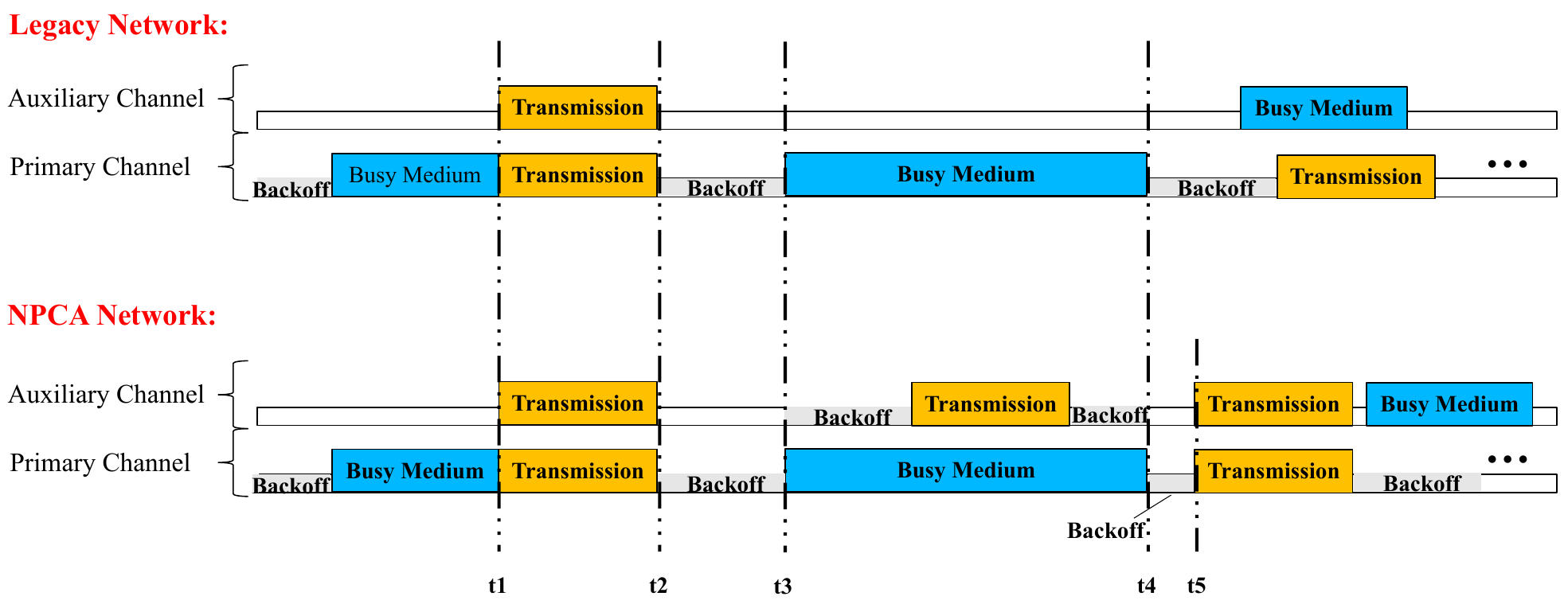}
    \caption{The Transmission Process Comparison between Legacy and NPCA Network.}
    \label{fig:LegacyvsNPCA}
\end{figure*}

Traditionally, as shown in Fig. \ref{fig:LegacyvsNPCA}, both types of devices are allowed to combine the primary channel and secondary channels for larger bandwidth transmission. However, this mechanism only works when the primary channel is idle in the legacy network, which results in
inefficient bandwidth utilization. The NPCA network is thereby used to alleviate such inefficiency. As depicted in Fig. \ref{fig:LegacyvsNPCA}, devices are allowed to check and switch to other available channels when the primary one is busy. These secondary channels are then used to continue the backoff process and transmit data. As shown in Fig. \ref{fig:channel},  the NPCA network first ranks and senses all non-primary channels. Then, the primary channel will be switched to the idle non-primary channel with the highest priority for transmission if the primary channel is busy. After the transmission, devices will switch back to the primary channel if it is sensed idle. For instance, in Fig. \ref{fig:LegacyvsNPCA}, initially, the NPCA and legacy network start transmission at the same time at $t_1$. However, after $t_3$, the legacy network stops the backoff process while the NPCA network switches to the non-primary channel and continues to backoff and transmit. After $t_4$, when sensing the primary channel is idle, the NPCA network switches back to the primary channel, and transmits one more packet at $t_5$. This example shows that NPCA network can transmit more packets than legacy network does during the whole transmission process. This approach helps to improve the overall throughput by making sure that available channels are utilized effectively.

% \begin{figure}[ht]
% \centering
%  \subfigure[The Backoff Process in Legacy Network.]
% {\includegraphics[width=0.48\textwidth]{figures/legacy.pdf}}
% \label{fig:2a}
% \centering \\
% \subfigure[The Backoff Process in NPCA Network.]{
% \includegraphics[width=0.48\textwidth]{figures/npca.pdf}}
% \label{fig:2b}
% \caption{The Transmission Process Comparison between Legacy and NPCA Network.}
% \label{fig:LegacyvsNPCA}
% \end{figure}

% \begin{figure}[htbp]
%     \centering
%     \subfigure[{\textwidth}]
%         \includegraphics[width=\textwidth]{figures/legacy.pdf}
%         \caption{The Backoff Process for Legacy Network}
%         \label{fig:sub1}
%     \\ % Change line
%     \subfigure[{\textwidth}]
%         \includegraphics[width=\textwidth]{figures/npca.pdf}
%         \caption{The Backoff Process for NPCA Network}
%         \label{fig:sub2}
%     \caption{The Comparison on Transmission Process for Legacy and NPCA Network}
%     \label{fig:test}
% \end{figure}

% If multiple non-primary channels are considered, the model also ranks them and selects the highest priority among all free channels when switching from the primary channel.

% \subsection{Model Throughput Analytics}

\section{System model}
\label{sec:sys_model}
Consider a scenario in a Base Station Set (BSS) where $n$ nodes are vying for a chance to transmit. We delve into the throughput analysis of this setting, primarily focusing on a single channel. Following Bianchi's model \cite{bianchi2000performance}, the throughput analyzed for single channel network, denoted as $S$, is a key performance metric. The computation of $S$ begins with determining $P_{tr}$, which is the probability of at least one node transmitting in a given slot time:
\begin{equation}
    P_{tr} = 1 - (1 - \tau)^{n}, \label{eq:ptr}
\end{equation}
where $\tau$ represents the likelihood of a station deciding to transmit during a random slot.

Denote $P_s$ as the probability of a successful transmission on the channel, which is given by
\begin{equation}
  P_{s} = \frac{n\tau(1-\tau)^{n-1}}{P_{tr}}, \label{eq:ps}
\end{equation}
which balances the chances of transmission against the odds of success.

Next, we denote $T_s$ as the average time the channel appears busy due to a successful transmission and $T_c$ as the average time it's busy during a collision. They are given by
\begin{flalign}
  T_{s} & = H + E[Pkt] + \text{SIFS} + \delta + \text{ACK} + \text{DIFS} + \delta, \label{eq:ts}&&\\
  T_{c} & = H + E[Pkt] + \delta + \text{EIFS}, \label{eq:tstc}&&
\end{flalign}
where $H$ is the PHY header, and $\delta$ is the propagation delay (typically 0.1 $\mu s$). $\text{SIFS}$, $\text{DIFS}$, and $\text{EIFS}$ (equal to $\text{SIFS} + \text{NACK} + \text{DIFS}$) are time intervals used for processing and responding to frames.

As a result, the average throughput, $S$ is expressed as
\begin{equation}
  S = \frac{{P_s}{P_{tr}}{E[P]}}{(1 - P_{tr})\sigma + {P_{tr}}{P_s}{T_s} + {P_{tr}}(1-{P_s}){T_c}}, \label{eq:s}
\end{equation}
where $\sigma$ represents the duration of an empty slot time, and $E[P]$ is the average packet payload size.

\begin{figure}[htp]
    \centering
\includegraphics[width=.2\textwidth]{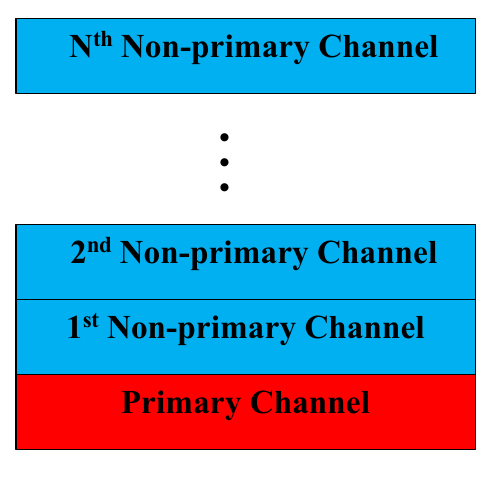}
    \caption{The Multi-channel Scenario.}
    \label{fig:channel}
\end{figure}

We further extend the analytical throughput model from the single channel to multiple channels, based on the throughput $S$ in Eq. (\ref{eq:s}). As Fig. \ref{fig:channel} shows, we consider a multi-channel scenario that includes one primary channel and $N$ non-primary channels. For the primary channel, the probability that it is sensed idle each time is $P_r$. For non-primary channels, the probability that the $n$th ($1\leq n \leq N$) channel is sensed idle is $P_n$. 

We first analyze the throughput of the legacy network for comparison. In a legacy network, non-primary channels are utilized only when the primary channel is idle and the device is ready to send packets. That is, If $k$ ($k \leq N$) channels are enabled to send packets simultaneously, the primary channel and the channels from the $1^{st}$ to the $(k - 1)^{th}$ must be idle, while the $k^{th}$ channel is busy. Under this scenario, the throughput for $k$ channels, denoted as $TH_k$, is given by
\begin{equation}
  TH_k = kS(P_r){P_1}{P_2} \cdots P_{k - 1} (1 - P_{k}), \label{eq:th_k}
\end{equation}
where $S(P_r)$ represents the single-channel throughput as calculated in Bianchi's model. When considering the scenario where all the channels (a total of $N + 1$) are sending packets, the throughput, $TH_{N+1}$, is formulated as
\begin{equation}
  TH_{N+1} = (N+1)S(P_r){P_1}{P_2} \cdots P_{N - 1}{P_N}. \label{eq:th_n}
\end{equation}
By considering all these scenarios, we obtain the throughput in the legacy network, $S_{leg}$, as
{\small\begin{equation}
  S_{leg} = \sum_{k=1}^{N+1}TH_k = S(P_r)\left(1 + \sum_{i=1}^{N}\prod_{j=1}^{i}P_j\right), \label{eq:s_leg}
\end{equation}}
where $P_j$ denotes the idle probability of the $j^{th}$ non-primary channel. For further analysis, we define the function $F(m)$ as
\begin{equation}
  F(m) = 1 + \sum_{i=m}^{N}\prod_{j=m}^{i}P_j, \label{eq:fm}
\end{equation}
where $m \leq N$ and both $i$ and $j$ start from $m$. This function represents the throughput coefficient assuming that the $m^{th}$ non-primary channel is the second priority channel, meaning all channels with a higher priority are busy. For instance, if $m = 1$, the $1^{st}$ non-primary channel is considered the second highest priority at that time, and in this case, $S_{leg} = S(P_r)F(1)$.

We then analyze the throughput of the NPCA network. We still consider the throughput aggregated across individual channels. First, the throughput on the primary channel, denoted as $TH_{pr}$, is given by
\begin{equation}
  S_{pr} = S(P_r)F(1), \label{eq:s_pr}
\end{equation}
which implies $TH_{pr} = S_{leg}$. When the primary channel is busy, the throughput contributed by other BSSs on the primary channel, $TH_{pr}^*$, is calculated with the idle probability $P_r$ as
\begin{equation}
  TH_{pr}^* = S(P_r)\left(\frac{1 - P_r}{P_r}\right). \label{eq:thpr_star}
\end{equation}
Since the probability that the $1^{st}$ non-primary channel is idle equals $P_1$, the corresponding throughput $TH_{1}$ should be
\begin{equation}
  TH_{1} = S(P_r)\left(\frac{1 - P_r}{P_r}\right)P_1, \label{eq:th1}
\end{equation}
 which also indicates $TH_{1} = TH_{pr}^*{P_1}$. Thus, the overall throughput when the $1^{st}$ non-primary channel is selected as the highest priority channel, $S_1$, is
\begin{equation}
  S_1 = S(P_r)\left(\frac{1 - P_r}{P_r}\right)P_1F(2). \label{eq:s1}
\end{equation}
Similarly, if we define the $S_n$ as the throughput on $n$th non-primary channel, we can infer that
\begin{equation}
\begin{aligned}
&S_2 = S(P_r)\left(\frac{1 - P_r}{P_r}\right)\left(\frac{1 - P_1}{P_1}\right)P_2 F(3), \\
&S_3 = S(P_r)\left(\frac{1 - P_r}{P_r}\right)\left(\frac{1 - P_1}{P_1}\right)\left(\frac{1 - P_2}{P_2}\right)P_3 F(4), \\
& \vdots \\
&S_{N-1} = S(P_r)\left(\frac{1 - P_r}{P_r}\right)\prod_{t=1}^{N-1}\left(\frac{1 - P_t}{P_t}\right)P_{N-1} F(N), \\
&S_{N} = S(P_r)\left(\frac{1 - P_r}{P_r}\right)\prod_{t=1}^{N}\left(\frac{1 - P_t}{P_t}\right)P_{N} F(N+1).
\end{aligned}
\end{equation}
Note that for $S_{N}$, $F(N+1) = 1$, indicating that there is no $(N+1)^{th}$ non-primary channel, so the coefficient is 1, considering only the scenario where the $N^{th}$ non-primary channel is transmitting. Taking all these scenarios into account, the throughput in NPCA network, $S_{npca}$, is derived as
 {\small\begin{equation}
\begin{aligned}
S_{npca} &= S_{pr} + \sum_{l=1}^{N}S_l \\
&= S(P_r)\left[1 + \sum_{i=1}^{N}\prod_{j=1}^{i}P_j + \frac{1 - P_r}{P_r}\sum_{t=1}^{N}P_t\left(1 + \sum_{i=t+1}^{N}\prod_{j=t+1}^{i}P_j\right)\right]. \label{eq:final_npca}
\end{aligned}
\end{equation}}
\begin{remark}
According to (\ref{eq:thpr_star}) and (\ref{eq:final_npca}), we can derive that $S_{leg} = S(P_r)\left(1 + \sum_{i=1}^{N}\prod_{j=1}^{i}P_j\right)$ while $S_{npca} = S_{leg} + S(P_r)\left[\frac{1 - P_r}{P_r}\sum_{t=1}^{N}P_t\left(1 + \sum_{i=t+1}^{N}\prod_{j=t+1}^{i}P_j\right)\right]$. As $S(P_r) > 0$,  $0 < P_r, P_t, P_j < 1$, 
demonstrating that $S_{npca} > S_{leg}$ always holds. That is, NPCA outperforms legacy network in terms of throughput.
\end{remark}

Meanwhile, we also analyze the access delay for the NPCA network. The access delay is defined as the time interval from the instant that a data packet is ready for transmission by a device to the instant that it begins transmission on the network medium. Drawing on the research by Taka Sakurai et al.\cite{4202176}, we find that with a large maximum contention window, the expected access delay $E[D]$ can be expressed as

{\small\begin{equation}
\begin{aligned}
  E[D] \approx &\frac{(t_{slot} + P_{tr}E[C^{*}])}{\tau(1 - P_{tr})} + \frac{(n - 1)(E[T^*] - E[C^*])}{1 - \tau} \\ 
  &+ \frac{P_{tr}E[C]}{(1 - P_{tr})} + E[T], \label{eq:access_delay}
\end{aligned}
\end{equation}}
where $t_{slot}$ denotes the slot duration, and $T$, a random variable (r.v.), represents the channel occupancy during a successful transmission by the tagged station. $C$ is a r.v. denoting the channel occupancy during a collision involving the tagged station, whereas $T^*$ and $C^*$ are r.v.s representing the channel occupancy for the corresponding phenomena but excluding the tagged station. Specifically,
\begin{equation}
\begin{aligned}
  T &= E[\text{Pkt}] + \text{DIFS}, \\
  C &= T^* = C^* = E[\text{Pkt}] + \text{SIFS} + \text{ACK} + \text{DIFS}. \label{eq:TC}
\end{aligned}
\end{equation}

% Employing this analytical model allows for the comparison of access delays in NPCA and legacy networks, facilitating further research into their interrelationships.

%\input{DRL}
\section{Simulation}
\label{simulation}

This section details our experiments on network performance, specifically focusing on throughput and packet delay in NPCA BSS and Legacy BSS systems. Initially, we examine a two-channel BSS under varying channel occupancy rates for both NPCA and legacy networks. We then explore a scenario where two BSS systems compete for transmission, analyzing their throughput and access delay. This helps us understand the interaction between NPCA and legacy networks when combined. Finally, we assess the delay in a two-BSS system across different network types. Table \ref{tb:simu_para} outlines the simulation parameters and channel transmission settings.

% \begin{figure}[tp]
%     \centering
% \includegraphics[width=.2\textwidth]{figures/Setup.pdf}
%     \caption{Setup of Simulation on two BSS.}
%     \label{fig:setup}
% \end{figure}

\begin{figure}[t]
\begin{minipage}[t]{0.48\linewidth}
\centering
 \subfigure[Single BSS scenario.]
{\includegraphics[width=1.2in]{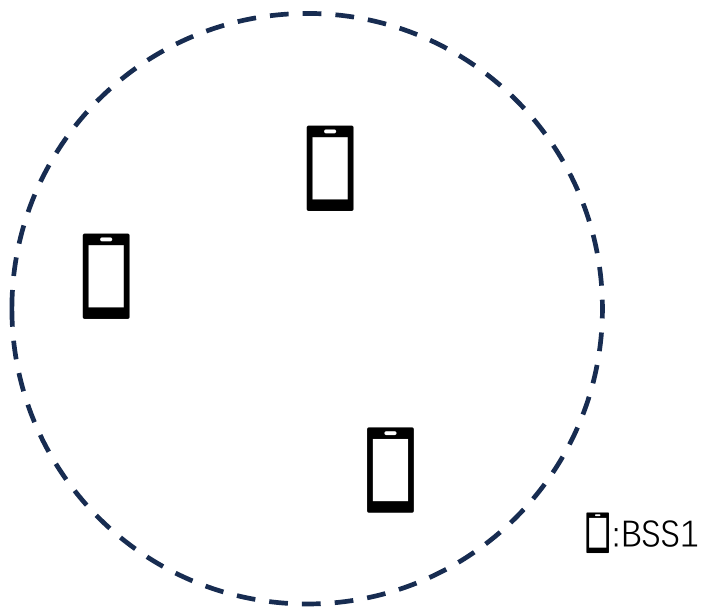}}
\label{fig:4a}
\end{minipage}
\begin{minipage}[t]{0.24\linewidth}
\centering
\subfigure[Two-BSS scenario.]{
\includegraphics[width=1.2in]{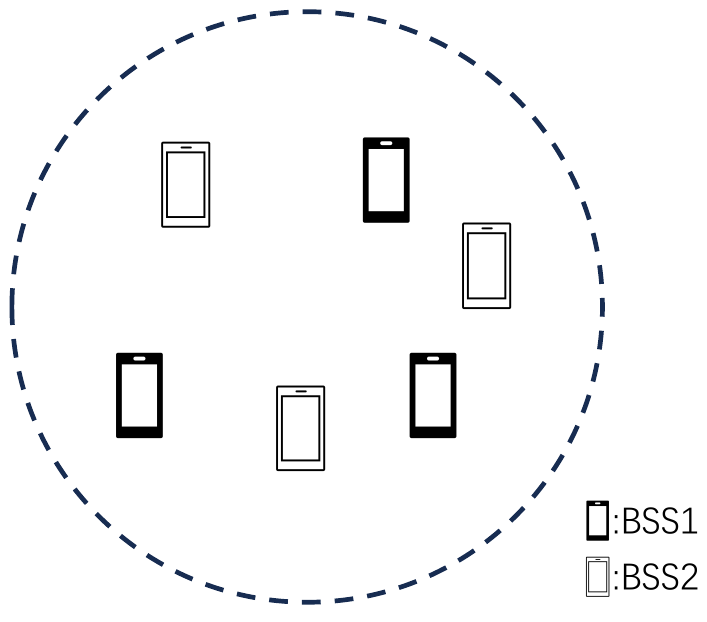}}
\label{fig:4b}
\end{minipage}%
\caption{Simulation Scenarios.}
\label{fig:Sim_Scenario}
\end{figure}

\begin{table}[htp]
\caption{Simulation Parameters}
\label{tb:simu_para}
\centering
\fontsize{8}{8}\selectfont{
\begin{tabular}{|c|c|}
\hline
\textbf{Parameters}         & \textbf{Value}  \\ \hline
Simulation Time (s) & 30 \\ \hline
Number of Channels & 2 \\ \hline
Primary Channel & Channel 1 \\ \hline
Channel Bandwidth (MHz) & 20 \\ \hline
Channel Utilization & 1 \\ \hline
Packet Size (Bytes) & 1500 \\ \hline
MCS  & 7 \\ \hline
CW Min, $CW_{min}$ & 16 \\ \hline
CW Max, $CW_{max}$ & 1024 \\ \hline
Slot (us) & 9 \\ \hline
SIFS (us) & 16 \\ \hline
DIFS (us) & SIFS + 2 $\cdot$ Slot \\ \hline
\end{tabular}
}
\end{table}

\subsection{Single BSS on Different Channel Occupancy Rate}
\begin{figure}[tp]
    \centering
\includegraphics[width=.48\textwidth]{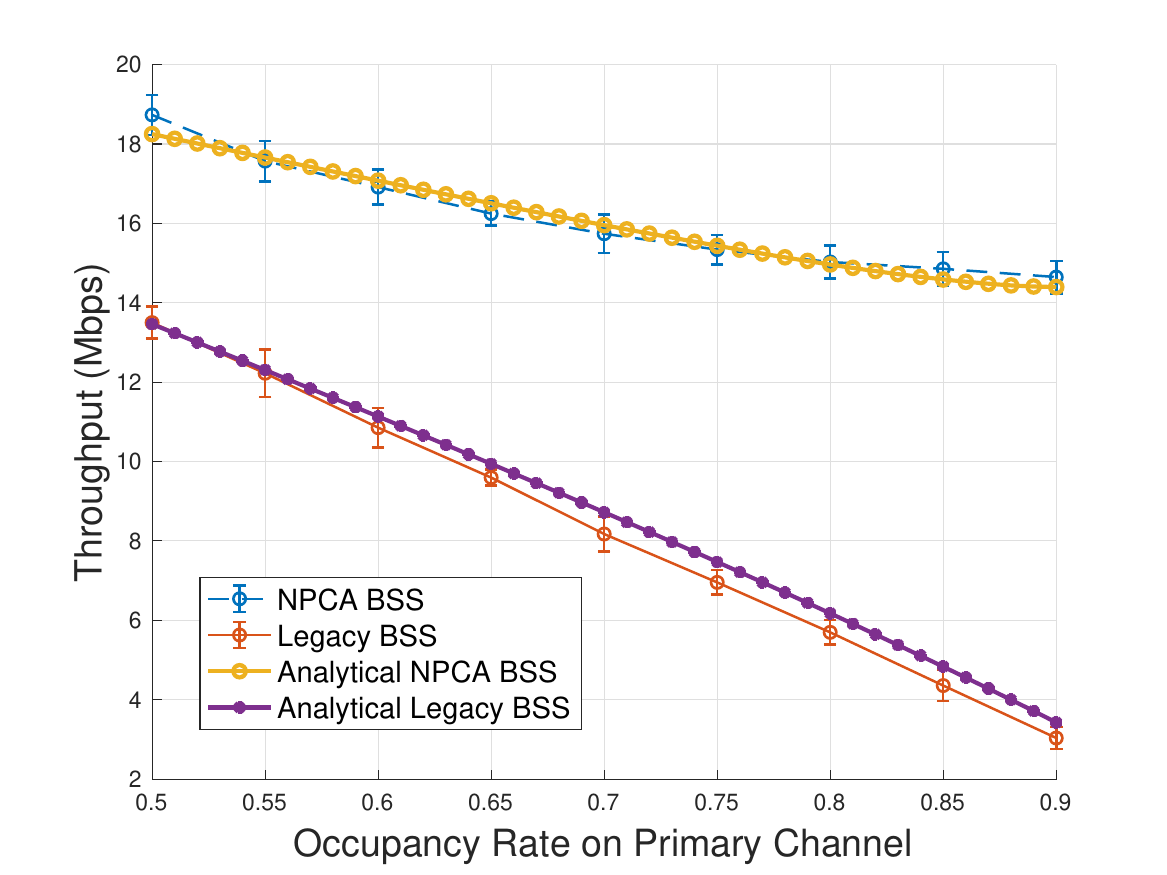}
    \caption{Single-BSS Network Throughput.}
    \label{fig:busy}
\end{figure}

We first explore how NPCA's throughput compares to the legacy network, particularly when the primary channel is highly occupied. We simulate a scenario where all stations within a single BSS can detect each other, as shown in Fig. \ref{fig:Sim_Scenario}(a). The legacy BSS, based on the Bianchi model, and NPCA model are simulated under varying primary channel occupancies. In this case, each BSS contains 10 stations, with Channel 1 designated as the primary channel. We simulate low occupancy (Idle rate: $P_{ch2} = 80\%$) on Channel 2 and very high occupancy (Idle rate: $P_{ch1} \leq 50\%$) on the primary channel (Channel 1) to observe the impact on network throughput. As depicted in Fig. \ref{fig:busy}, where it illustrates the throughput of 40 MHz networks when the Non-Primary Channel is relatively idle ($P_{ch2} = 80\%$) while Primary Channel is busy ($P_{ch1} \leq 50\%$), that the legacy network struggles with high occupancy on the primary channel, leading to more collisions and waiting periods due to Bianchi's rule. Conversely, NPCA demonstrates significantly better throughput, particularly when the primary channel is busy. NPCA achieves this by switching to the secondary channel (Channel 2) when it senses congestion on the primary channel. The busier the primary channel, the more beneficial it is to switch to Channel 2, showcasing NPCA's advantage in utilizing secondary channels to enhance network quality.

% \begin{table}[htp]
% \caption{Setups for One-BSS Throughput Analysis}
% \label{tb:one-bss}
% \centering
% \fontsize{8.5}{12}\selectfont{
% \begin{tabular}{|c|c|}
% \hline
% \textbf{Parameters}         & \textbf{Value}  \\ \hline
% Number of Stations & 10 \\ \hline
% Channel Numbers & 2 \\ \hline
% Primary Channel & Channel 1 \\ \hline
% Channel Bandwidth (MHz) & 20 \\ \hline
% Channel 1 Idle Rate (\%), $P_{ch1}$ & $\leq$ 50 \\ \hline
% Channel 2 Idle Rate (\%), $P_{ch2}$ & 80 \\ \hline
% Channel Utilization & 1 \\ \hline
% \end{tabular}
% }
% \end{table}

\begin{figure}[htp]
    \centering
\includegraphics[width=.48\textwidth]{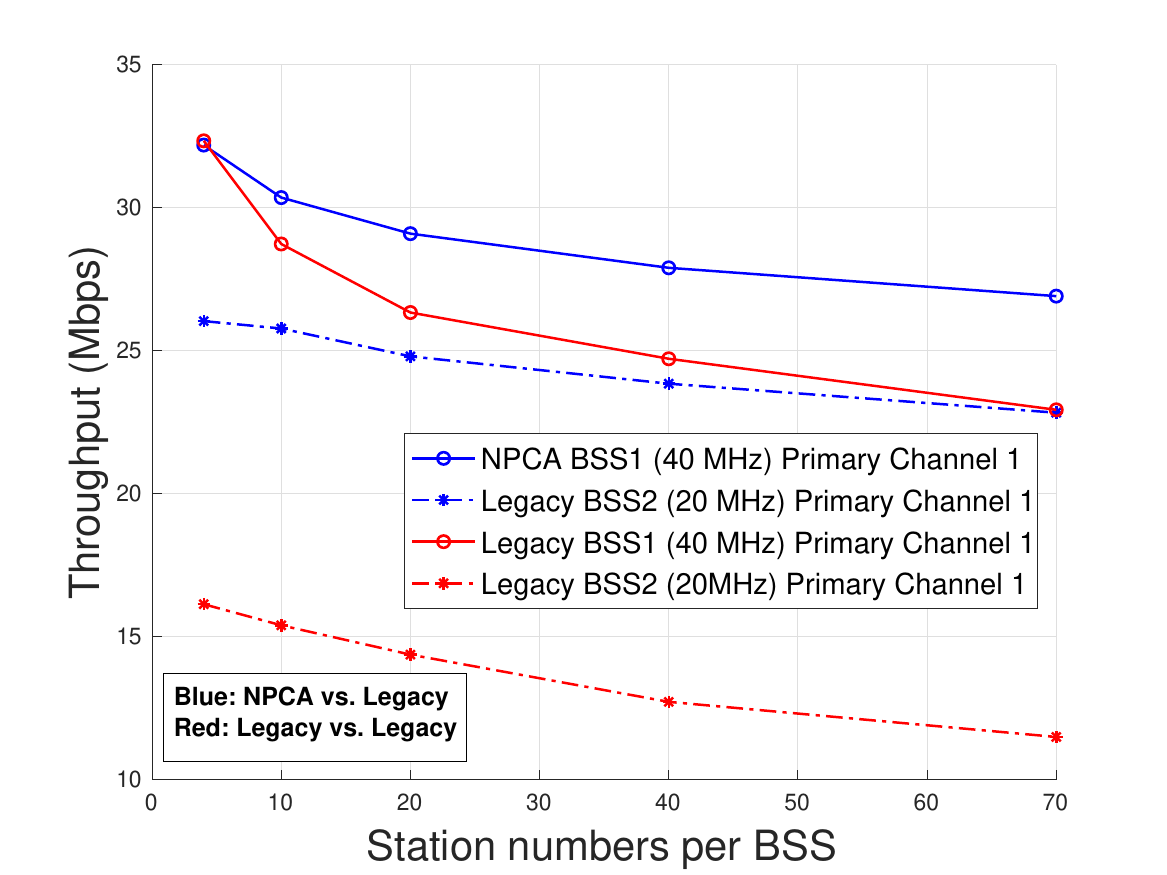}
    \caption{Two-BSS Network Throughput.}
    \label{fig:nl_ll}
\end{figure}

\subsection{Two-BSS System}
We conducted two simulations for this model, involving one 40 MHz (two-channel) and one 20 MHz (single-channel) BSS within the same sensing area. As depicted in Fig. \ref{fig:Sim_Scenario}(b), each BSS has an equal number of stations, and all the nodes can sense each other. Intriguingly, as shown in Fig. \ref{fig:nl_ll}, it indicates the comparison of two-BSS system's performance in two scenarios. In the first scenario (red lines), both BSS systems are legacy networks. The simulated throughput in Fig.~\ref{fig:nl_ll} matches our analytical model, and we observe that the throughput of the 40 MHz BSS (BSS1) is twice as much as the throughput of 20 MHz BSS (BSS2), attributed to its dual-channel capability allowing two packets to be transmitted simultaneously. As the number of stations increases, the likelihood of collisions also rises, leading to reduced throughput following Bianchi's rule. In the second scenario (blue lines), BSS1 (40MHz) becomes an NPCA network while BSS2 (20MHz) remains a legacy network. We can see that NPCA's throughput surpasses that of the legacy network under equal station counts. Additionally, when paired with the NPCA network, the legacy network's throughput also improves significantly. This enhancement occurs because BSS1, upon detecting the transmission from BSS2, opts for the secondary channel for backoff or transmission, thus increasing its transmission probability. Meanwhile, BSS2 gets a better chance to transmit on Channel 1.

\subsection{Access Delay Analysis}
\begin{figure}[tp]
    \centering
\includegraphics[width=.48\textwidth]{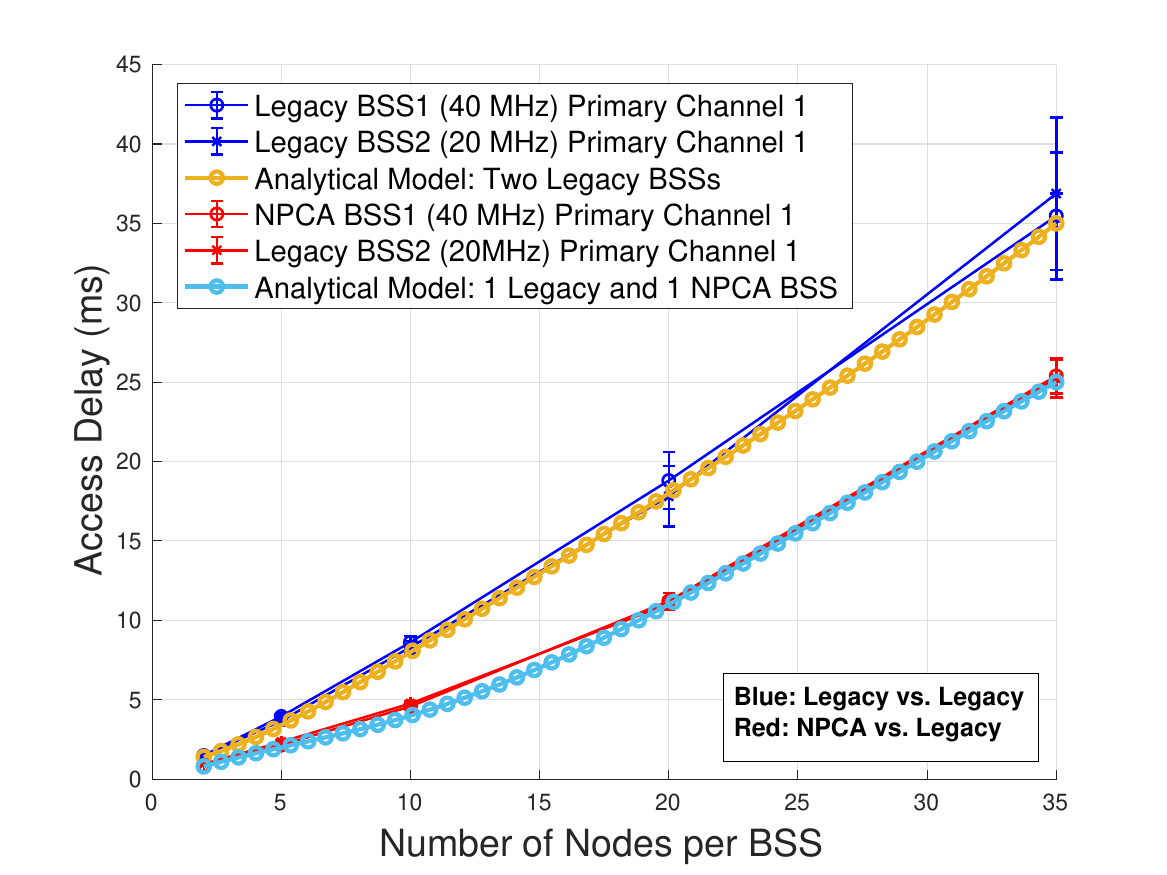}
    \caption{Two-BSS Network Average Delay.}
    \label{fig:delay}
\end{figure}

As illustrated in Fig. \ref{fig:delay}, we expand upon the two-BSS model simulation to investigate its access delay. Access delay is the total time a data packet takes to move from its starting point to its destination. This includes all types of delays, such as those caused by the data waiting in line (queuing), the time it takes to process the data at each stop along the way (processing), the physical time it takes to move the data (propagation), and the time to actually send the data (transmission delays). Understanding access delay is key for evaluating how well different network protocols and devices perform under a variety of conditions. This knowledge helps in designing and optimizing networks to ensure they deliver data quickly and reliably, which is particularly important for applications that need fast responses.

In our simulations, we observed that having more nodes in the BSS tends to increase the likelihood of collisions. This means data packets have to be sent again (retransmitted) and wait longer to get through. However, the NPCA model helps reduce access delay, making the network not only quicker but also more consistent in its performance. This stability is crucial for a network's efficiency. The reason behind this improvement is twofold. First, NPCA is better at managing how channels are used. It can move transmissions to other channels when necessary, making sure the network keeps running smoothly without bottlenecks. Second, it reduces the number of collisions, which means data flows more freely without as many interruptions. This leads to a network not only faster but also more reliable, with less variation in delay times, enhancing overall network performance.

\section{Conclusion}
\label{sec.conclusion}
This paper presented a comprehensive study on the NPCA protocol proposed by the IEEE 802.11 UHR group. We developed a novel analytical model for the NPCA network, which was further validated by the simulation. We showed that the NPCA network outperforms the legacy network in terms of throughput by optimizing the utilization of available bandwidth across primary and secondary channels, especially under the busy primary channel. Meanwhile, we also evaluated the performance of the NPCA network in coexistence with the legacy network, revealing that the NPCA also impacts the legacy network by not only significantly improving the throughput but also maintaining a lower level of access delay. Our contribution to the ongoing discussions within the IEEE 802.11 UHR group provides analytical insights regarding the NPCA implementation in the upcoming 802.11bn.

%\input{conclusion}

% \appendix

% \input{appendix}

\bibliographystyle{IEEEtran}
\bibliography{reference}
\end{document}